\documentclass[aps,pra,reprint,superscriptaddress,amsmath,amssymb]{revtex4-2}
\usepackage{graphicx}
\usepackage{xcolor,mathrsfs}
\usepackage[colorlinks=true,bookmarks=false,citecolor=blue,urlcolor=blue]{hyperref} %pdflatex
\usepackage{float}
\usepackage{braket}

\begin{document}

\title{Enantio-selective inverse Faraday effect in isotropic chiral molecular mixtures}

\author{Raju Adhikary} 
\email{raju.adhikary@graduate.univaq.it}
\affiliation{Department of Physical and Chemical Sciences, University of L'Aquila, Via Vetoio, 67100 L'Aquila, Italy}

\author{Ambaresh Sahoo} 
\affiliation{Department of Physical and Chemical Sciences, University of L'Aquila, Via Vetoio, 67100 L'Aquila, Italy}

\author{Matteo Silvestri} 
\affiliation{Department of Physical and Chemical Sciences, University of L'Aquila, Via Vetoio, 67100 L'Aquila, Italy}

\author{Massimiliano Aschi} 
\affiliation{Department of Physical and Chemical Sciences, University of L'Aquila, Via Vetoio, 67100 L'Aquila, Italy}

\author{Antonio Mecozzi} 
\affiliation{Department of Physical and Chemical Sciences, University of L'Aquila, Via Vetoio, 67100 L'Aquila, Italy}

\author{Davide Tedeschi} 
\affiliation{Department of Physical and Chemical Sciences, University of L'Aquila, Via Vetoio, 67100 L'Aquila, Italy}

\author{Carino Ferrante} 
\affiliation{CNR-SPIN, c/o Dip.to di Scienze Fisiche e Chimiche, Via Vetoio,  L'Aquila 67100, Italy}

\author{Andrea Marini} 
\email{andrea.marini@univaq.it}
\affiliation{Department of Physical and Chemical Sciences, University of L'Aquila, Via Vetoio, 67100 L'Aquila, Italy}
\affiliation{CNR-SPIN, c/o Dip.to di Scienze Fisiche e Chimiche, Via Vetoio,  L'Aquila 67100, Italy}

\begin{abstract}
Enantiomeric excess detection in a chiral molecular mixture is paramount because very often opposite enantiomers exhibit profound functional dissimilarities that play decisive roles in biochemical applications. Existing chiral sensing methods mostly rely on large operational sample volumes, hindering compatibility with integrated sensing schemes. Here, we propose a novel chiroptical sensing technique based on the inverse Faraday effect in a photonic micro-capillary filled with nl-volume chiral drug solution. We theoretically demonstrate that, upon excitation by intense laser light, an isotropic assembly of chiral drugs produces a static magnetisation, with amplitude and direction depending on the enantiomeric excess. In turn, by measuring the chirally-sensitive static magnetic field in the vicinity of the micro-tube one can retrieve the enantiomeric excess of the chiral drug solution. Our theoretical predictions unlock new opportunities for the development of innovative nanophotonic devices suitable for efficient chiroptical sensing with nl-volume sensitivity.
\end{abstract}

\maketitle

\textit{Introduction}--Chirality, i.e., asymmetry upon mirror reflection, is omnipresent at both microscopic and macroscopic scales in nature. In particular, chiral molecular sensing holds significant implications in drug toxicity and functionality assessment across biological and pharmaceutical applications \cite{Nguyen2006,Blackmond2010, McVicker2024}. For many years, optics-based enantiomeric excess detection has exploited chiroptical interactions between polarised light and chiral molecules, producing rotatory dispersion, electronic/vibrational circular dichroism, and Raman optical activity \cite{Muller2000, Polavarapu2007, Wesolowski2013}. Owing to the intrinsically weak molecular chiroptical interaction, such so-far-established spectroscopic techniques are suitable for the analysis of macroscopic chiral bulk samples with volumes in the ml range, while they are inadequate for lab-on-chip operation. 

Currently, novel promising chiroptical sensing platforms based on superchirality \cite{Tang2010, Pellegrini2017} and nanophotonics\cite{Solomon2020}, e.g., enhanced light-matter interactions facilitated by plasmonic nanostructures \cite{Fan2010, Govorov2011, Slocik2011, Hentschel2012, Nesterov2016, Neubrech2017, Venturi2023, Adhikary2025} and metasurfaces \cite{Mohammadi2018, Solomon2018, Mohammadi2019, Mattioli2020, Miao2021}, are being developed. Such advanced nanophotonic architectures enable great control over light manipulation at subwavelength scales, where interaction between chiral nanostructures and polarised light is enhanced \cite{Olohan2026, Jones2024}, unlocking previously unattained magneto-optical applications, e.g., through inverse Faraday effect \cite{Mou2023, Hareau2025}. Furthermore, in spite of the even weaker nonlinear chiroptical interactions \cite{Fischer2005}, innovative ultrafast chiral sensing schemes enable chiral dynamics probing \cite{Ayuso2021, Ayuso2022, Wanie2024} and the exploitation of synthetic chiral light fields for molecular fingerprinting \cite{Mayer2024}. Nevertheless, the development of efficient nanophotonic platforms for highly sensitive chiroptical sensing remains a major bottleneck for practical implementations.

\begin{figure}[b]
	\centering
	\begin{center}	
    \includegraphics[width=0.48\textwidth ]{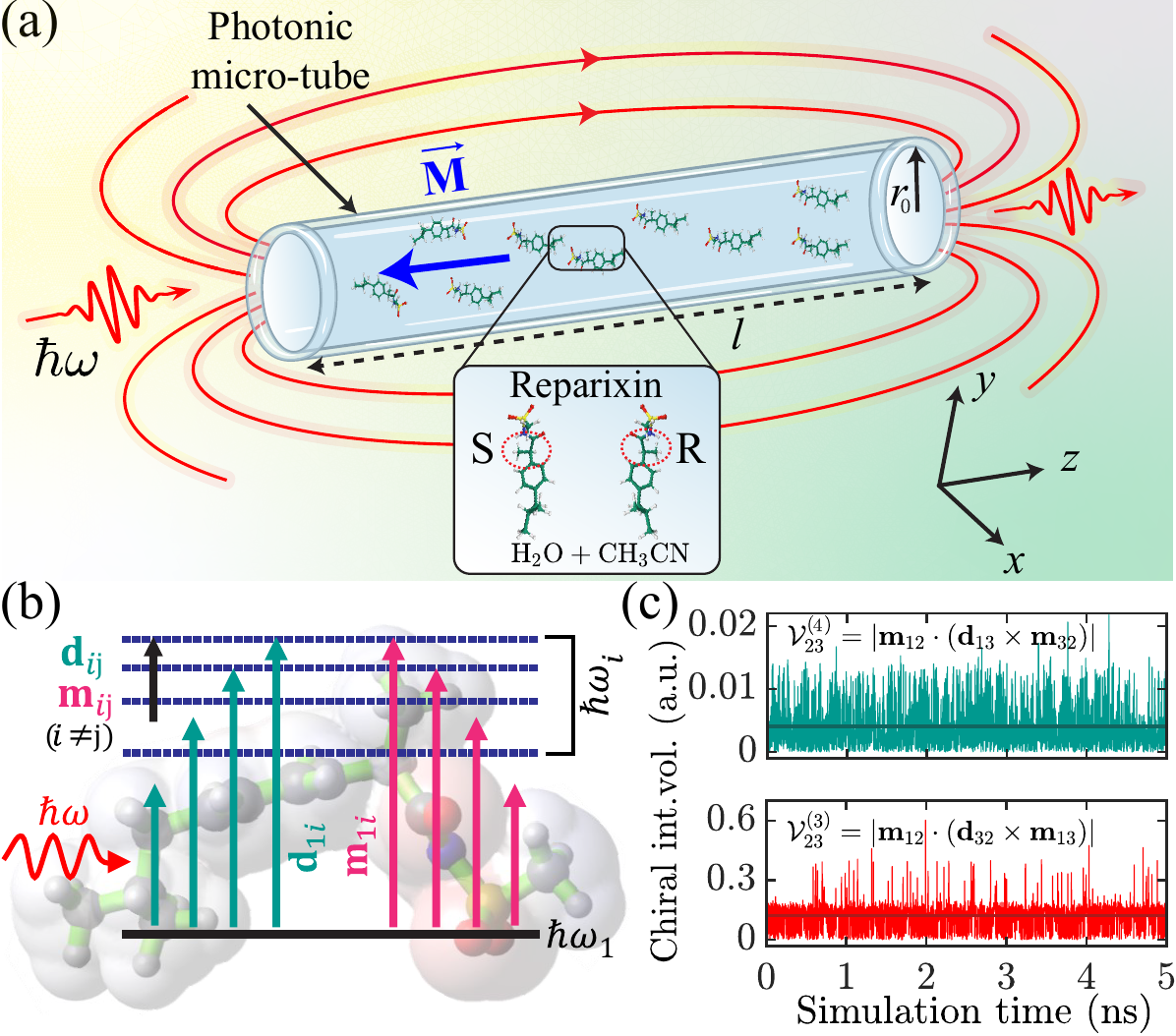}
		\caption{(\textbf{a}) Schematic illustration of the considered PMC with length $l=1~{\rm mm}$ and radius $r_0=50~\mu{\rm m}$, embedding a CDS of reparixin dissolved in a 50:50 MS of water and acetonitrile. (\textbf{b}) Schematic illustration of the electronic structure and the relevant electronic transitions of reparixin when illuminated by a laser field within the electric/magnetic dipole approximation. (\textbf{c}) Temporal evolution of relevant chiral interaction volumes $|\mathcal{V}_{23}^{(1)}|$ and $|\mathcal{V}_{23}^{(2)}|$ as a function of MD-simulation frame (expressed in nanoseconds) corresponding to a single reparixin molecule in the $k=1$ conformational state, obtained through MD-PMM calculations adopting a cubic simulation box of volume $\simeq 27~{\rm nm}^{3}$.} 
		\label{Fig1}
	\end{center}
\end{figure}

In this Letter, we develop an innovative approach for enantiomeric discrimination of nl-volume chiral drug solutions through inverse Faraday effect (IFE) in a photonic micro-capillary (PMC) embedding isotropic chiral molecular solutions, see Fig.~\ref{Fig1}({\bf a}). IFE is a nonlinear (NL) magneto-optical phenomenon whereby circularly polarised (CP) light induces an effective quasi-static magnetisation (QSM) in matter, with direction controlled by the optical spin angular momentum. While IFE has been recently demonstrated by ordered nanophotonic structures for magneto-optical applications \cite{Mou2023, Hareau2025}, 
for the first time to our knowledge, we here theoretically demonstrate that isotropic assemblies of randomly oriented chiral molecules produce enantio-selective IFE. We observe that, upon excitation by an intense optical pulse, the chiral mixture produces a QSM parallel to the PMC due to such a second-order NL effect. We find that such a QSM possesses both chirally-insensitive and enantio-selective contributions, generating a quasi-static magnetic field (QSMF) that encodes the chiral mixture enantiomeric excess. Thus, by measuring such a QSMF, one can retrieve the enantiomeric excess after calibration and data post-processing. Our results are obtained for a realistic chiral drug solution (CDS), reparixin (a therapeutic agent employed for the treatment of COVID-19 pneumonia \cite{Landoni2022}) dissolved in a mixed solvent (MS) of water and acetonitrile with 50:50 volume ratio. Light-driven electron dynamics in such a molecule is modelled in the density matrix formalism, incorporating microscopic quantum molecular observables (QMOs) obtained through robust quantum chemical computations. Considering a realistic PMC with length $l = 1~{\rm mm}$, radius $r_0 = 50~\mu{\rm m}$, and embedding the considered dilute CDS (with dilute molecular number density $n_{\rm mol}\simeq 10^{-1}-10^{-3}~{\rm nm}^{-3}$), we attain a QSM with amplitude of the order of $|{\bf M}_{\rm IFE}| \simeq 100$ ${\rm kA}/\mu{\rm m}$, which produces chirally-sensitive QSMF of the order $|{\bf B}_{\rm IFE}|\simeq 1$ ${\rm nT}$ order. Our findings reveal a novel chiroptical effect that holds great potential for the development of promising nanophotonic devices for efficient chiral sensing of nl-volume isotropic chiral drugs.

\textit{Microscopic electron dynamics}--We investigate the microscopic response of reparixin upon excitation by external radiation in the density matrix formalism, incorporating relevant QMOs, i.e., electronic energies, static and transition electric/magnetic dipole moments of the considered CDS. Such QMOs are evaluated as quantum expectation values and averaged statistical-mechanical quantities across a broad set of reparixin-MS configurations, i.e., time averages over MD-timescales ($\simeq {\rm ns}$) in the ergodic assumption. This is accomplished by employing previously reported results based on Molecular Dynamics (MD) simulations complemented with time-dependent Density Functional Theory (TD-DFT) and Perturbed Matrix Method (PMM) simulations \cite{Venturi2023}, see Supplementary Information (SI) for further details. The MD-trajectory analysis of solvated reparixin indicates four statistically relevant conformational states, labelled by the index $k=1-4$ at equilibrium \cite{Venturi2023}, which are considered to obtain ensemble-averaged QMOs in our calculations. The interactions of reparixin enantiomers (hereafter marked by the index ${\rm u = S, R}$) with external radiation electric ${\bf E}({\bf r},t)$ and magnetic ${\bf B}({\bf r},t)$ fields are accounted for by considering the total Hamiltonian $\hat{\mathcal{H}}_{\rm T}^{(k,\rm u)} = \sum_{i} \hbar \omega_{i}(k)|\Psi_i(k,{\rm u})\rangle \langle \Psi_i(k,{\rm u})| - {\bf E}({\bf r},t) \cdot \hat{{\bf d}}(k,{\rm u}) - {\bf B}({\bf r},t) \cdot \hat{{\bf m}}(k,{\rm u})$, where the index $i$ labels the $|\Psi_i(k,{\rm u})\rangle$ electronic eigenstate with energy eigenvalue $\hbar \omega_{i}(k)$, and the electric/magnetic dipole operators $\hat{{\bf d}}(k,{\rm u}),\hat{{\bf m}}(k,{\rm u})$ are expressed in the dyadic representation $\hat{{\bf a}}(k,{\rm u}) = \sum_{i,j} {\bf a}_{i,j}(k,{\rm u})|\Psi_j(k,{\rm u})\rangle \langle \Psi_i(k,{\rm u})|$ with ${\bf a} = {\bf d},{\bf m}$. In this study we consider $i=1-5$ electronic states leading to the lowest-energy convoluted UV-absorption peak in  previously reported MD-PMM calculations \cite{Venturi2023}. Thus, electrons undergo multilevel radiation-induced electronic transitions, both from the electronic ground $(i=1)$ state $\hbar \omega_1 \rightarrow \hbar \omega_i$ (where $i=2,3,4,5$ represents higher energy states) and excited states $\hbar \omega_i \rightarrow \hbar \omega_j$ (between higher energy states $i,j\geq2$ with $i\neq j$). Such electron dynamics are illustrated in Fig.~\ref{Fig1}({\bf b}), depicting relevant electric ${\bf d}_{1i},{\bf d}_{ij}$ and magnetic ${\bf m}_{1i},{\bf m}_{ij}$ transition dipole moments. Note that reparixin possesses vanishing static magnetic dipole moments (${\bf m}_{ii}=0$ for all states) attributed to the opposite-spin electron sets present in the ground state, while the static electric dipole moments ${\bf d}_{ii}$ are finite. While electric dipole moments are purely real polar vectors ${\bf d}_{i,j}(k,{\rm u}) \in \mathbb{R}^3$, magnetic dipole moments are purely imaginary axial vectors ${\bf m}_{i,j}(k,{\rm u}) \in \mathbb{I}^3$, and thus for opposite enantiomers they transform as ${\bf d}_{i,j}(k,{\rm S}) = \mathscr{R}_{\hat{\bf n}}{\bf d}_{i,j}(k,{\rm R})$ and ${\bf m}_{i,j}(k,{\rm S}) = -\mathscr{R}_{\hat{\bf n}}{\bf m}_{i,j}(k,{\rm R})$ upon the reflection operator $\mathscr{R}_{\hat{\bf n}}={\rm I}-2\hat{\bf n}\hat{\bf n}$, operarating across a broken-symmetry plane perpendicular to the unit vector $\hat{\bf n}$. Remarkably, such transformations allow us to observe that the quantities
\begin{align}\label{Eq1}
    &\mathcal{V}_{1,i}(k,{\rm u}) = {\bf m}_{1,i}(k,{\rm u}) \cdot[{\bf d}_{1,i}(k,{\rm u}) \times ({\bf d}_{1,1}(k,{\rm u}) - {\bf d}_{i,i}(k,{\rm u}))], \nonumber\\
    &\mathcal{V}_{i,j}^{(1)}(k,{\rm u}) = {\bf m}_{1,i}(k,{\rm u}) \cdot[{\bf d}_{1,j}(k,{\rm u}) \times {\bf d}_{j,i}(k,{\rm u})],\nonumber \\
    &\mathcal{V}_{i,j}^{(2)}(k,{\rm u}) = {\bf m}_{1,i}(k,{\rm u}) \cdot[{\bf m}_{j,i}(k,{\rm u}) \times {\bf m}_{1,j}(k,{\rm u})],
\end{align}
with $i,j=2 \rightarrow 5$, $j\neq i$, are pure scalars, while the triple products 
\begin{align}\label{Eq2}
&\mathcal{V}_{i,j}^{(3)}(k,{\rm u}) = {\bf m}_{1,i}(k,{\rm u}) \cdot[{\bf d}_{j,i}(k,{\rm u}) \times {\bf m}_{1,j}(k,{\rm u})] , \nonumber \\
&\mathcal{V}_{i,j}^{(4)}(k,{\rm u}) = {\bf m}_{1,i}(k,{\rm u}) \cdot[{\bf d}_{1,j}(k,{\rm u}) \times {\bf m}_{j,i}(k,{\rm u})],
\end{align}

\begin{figure*}[t!]
	\centering
	\begin{center}
		\includegraphics[width=\textwidth]{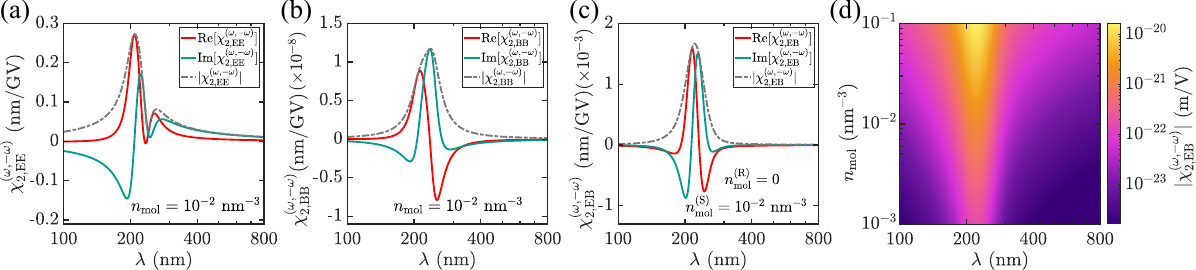}
		\caption{(\textbf{a}-\textbf{c}) Dependence of complex second-order NL susceptibilities (\textbf{a}) $\chi_{2,{\rm EE}}^{(\omega,-\omega)}$, (\textbf{b}) $\chi_{2,{\rm BB}}^{(\omega,-\omega)}$, and (\textbf{c}) $\chi_{2,{\rm EB}}^{(\omega,-\omega)}$ over the vacuum wavelength $\lambda $, calculated for pure-S CDS with molecular number density $n_{\rm mol}^{\rm (S)} = 10 ^{-2}~{\rm nm} ^{-3}$ (and $n_{\rm mol}^{\rm (R)} = 0$). (\textbf{d}) Color plot illustrating the dependence of  $|\chi_{2,{\rm EB}}^{(\omega,-\omega)}|$ over $\lambda$ and the total molecular number density $n_{\rm mol} = n_{\rm mol}^{\rm (S)} $, i.e., $\Delta n_{\rm mol} = n_{\rm mol}$ for pure-S CDS.} 
		\label{Fig2}
	\end{center}
\end{figure*}

\noindent are {\it enantio-selective pseudoscalars}, i.e., $\mathcal{V}_{i,j}^{(3,4)}(k,{\rm S})=-\mathcal{V}_{i,j}^{(3,4)}(k,{\rm R})$. These microscopic elements represent effective chiral interaction volumes subject to the orientation of three electric/magnetic dipole matrix elements, and in turn, they play decisive roles in second-order NL interactions, see below. We illustrate the MD-induced temporal evolution of  $\mathcal{V}_{i,j}^{(3)}$ and $\mathcal{V}_{i,j}^{(4)}$ in Fig.~\ref{Fig1}({\bf c}), evaluated for pure-S solvated reparixin in the conformation state $k = 1$.  For each conformation state $k$ and enantiomeric form ${\rm u= R, S}$, radiation-induced electron dynamics is governed by the density matrix equations
\begin{equation} \label{Eq3}
    d\hat{\rho}_{k,{\rm u}}/dt = (1/{i\hbar})[\hat{\mathcal{H}}_{\rm T}^{(k,\rm u)},\hat{\rho}_{k,{\rm u}} ] + \mathcal{L}(\hat{\rho}_{k,{\rm u}}), 
\end{equation}
where $\hat{\rho}_{k,{\rm u}}({\bf r},t)$ and $\mathcal{L}(\hat{\rho}_{k,{\rm u}}) $ indicate the time-dependent electronic density matrix and the Lindblad operator, respectively. For each electronic transition, Lindblad relaxation rates are determined from the absorption bandwidth obtained from previous MD-PMM calculations \cite{Venturi2023}, see SI for further detail.

\textit{Microscopic NL-IFE hyper-polarisabilities}--We perturbatively solve Eq. (\ref{Eq3}) up to second order to obtain $\hat{\rho}_{k,{\rm u}}({\bf r},t)$ in the slowly-varying envelope approximation (SVEA), i.e., by considering weak quasi-monochromatic external radiation fields ${\bf V}({\bf r},t) = {\rm Re}[{\bf V}_0({\bf r}, t)e^{-i\omega t}]$, with ${\bf V} = {\bf E,B} \propto \varepsilon_{\rm exp}$ ($\varepsilon_{\rm exp}<<1$ is a small dimensionless dummy variable) and carrier wavelength $\lambda = 2\pi c/\omega$, where $c$ is the light speed in vacuum and $\omega$ is the carrier angular frequency. In order to do so, we express the density matrix in the dyadic form $\hat{\rho}_{k,{\rm u}}=\sum_{i,j=1}^{5}\rho_{i,j}^{(k,{\rm u})}|\Psi_i(k,{\rm u})\rangle\langle\Psi_j(k,{\rm u})|$, where $\rho_{i,j}^{(k,{\rm u})}$ are time-dependent density-matrix elements. In the weak-excitation limit, we expand the density matrix elements as $\rho_{i,j}^{(k,{\rm u})}=\rho_{i,j}^{(0,k,{\rm u})}+\rho_{i,j}^{(1,k,{\rm u})}+\rho_{i,j}^{(2,k,{\rm u})}+\ldots$, where $|\rho_{i,j}^{(1,k,{\rm u})}|\propto \varepsilon_{\rm exp}$, and $|\rho_{i,j}^{(0,k,{\rm u})}|,|\rho_{i,j}^{(2,k,{\rm u})}|\propto \varepsilon_{\rm exp}^2$. By inserting this perturbative expansion into Eq. (\ref{Eq3}), we isolate the first- and the second-order density-matrix contributions, calculating $\rho_{i,j}^{(0,k,{\rm u})},\rho_{i,j}^{(1,k,{\rm u})}$, and $\rho_{i,j}^{(2,k,{\rm u})}$, see SI for further detail. At first order, electron dynamics produces previously reported bi-anisotropic linear response \cite{Venturi2023}. Second-order NL contributions produce diverse effects worth of investigation \cite{Ayuso2022bis}, but in the present study we retain only terms oscillating at zero carrier frequency, thus focusing solely on the IFE. Thus, for each  enantiomeric form ${\rm u=S,R}$, we calculate the expectation value of the IFE-induced magnetic dipole operator as
${\bf m}_{\rm IFE}^{({\rm u})} = \sum_{k=1}^{4}p_k{\rm Tr}[\hat{\rho}_{k,{\rm u}}^{(2)}\hat{\bf m}(k,{\rm u})]$, where $\hat{\rho}_{k,{\rm u}}^{(2)}$ denotes the zero-frequency second-order contribution to the density matrix. Note that ${\bf m}_{\rm IFE}^{({\rm u})}$ is conformationally averaged with probabilities $p_{k} \simeq 0.25$ in order to account for the statistically-relevant thermodynamic realisations $k=1-4$ \cite{Venturi2023}. The IFE-induced magnetic dipole is further averaged over arbitrary molecular rotations $\langle ... \rangle$ through the Euler rotation matrix approach \cite{Andrews2004}, providing 
\begin{equation} \label{Eq4}
    \begin{aligned}
        &\langle {\bf m}_{\rm IFE}^{(\rm u)} \rangle =\dfrac{1}{\mu_0c}{\rm Re} \Big[ \beta_{\rm EE}^{\rm (u)}(\omega) \left( {\bf E}_0 \times {\bf E}_0^* \right) +  \quad \quad \quad\\  
        &\ \ + c\beta_{\rm EB}^{\rm (u)} (\omega) \left( {\bf E}_0 \times {\bf B}_0^* \right) + c^2\beta_{\rm BB}^{\rm (u)}  (\omega) \left( {\bf B}_0 \times {\bf B}_0^* \right) \Big], 
    \end{aligned} 
\end{equation}
where $\mu_0$ is the vacuum permeability, and $\beta_{\rm EE/EB/BB}^{\rm (u)}$ are the NL-IFE isotropic hyper-polarisabilities of each enantiomeric form ${\rm u=S,R}$, which analytical expressions are provided in the SI. As anticipated above, these NL-IFE hyper-polarisabilities depend on triple products accounting for effective chiral interaction volumes. Note that, while $\beta_{\rm EE/BB}^{(\rm R)}(\omega) = \beta_{\rm EE/BB}^{(\rm S)}(\omega)$ are chirally-insensitive, $\beta_{\rm EB}^{(\rm R)}(\omega) = - \beta_{\rm EB}^{(\rm S)}(\omega)$ is an enantio-selective pseudoscalar.

\textit{Macroscopic NL-IFE response}--The CDS effective macroscopic NL response is evaluated by considering dilute molecular densities $n_{\rm mol} = n_{\rm mol}^{\rm (S)} + n_{\rm mol}^{\rm (R)} \approx 10^{-1} - 10^{-3}~{\rm nm}^{-3}$. In such a dilute-medium approximation, the macroscopic IFE-QSM can be evaluated as ${\bf M}_{\rm IFE} ({\bf r},t) = \sum_{\rm u =S, R}n_{\rm mol}^{\rm (u)} \langle {\bf m}_{\rm IFE}^{\rm (u)} \rangle$, providing 
\begin{equation} \label{Eq5}
    \begin{aligned}    
    &{\bf M}_{\rm IFE} ({\bf r},t) = \dfrac{1}{\mu_0c}{\rm Re}\Big[ \chi_{2, \rm EE}^{(\omega,-\omega)}({\bf E}_0 \times {\bf E}_0^*) +\\
    &+c\chi_{2, \rm  EB}^{(\omega,-\omega)}({\bf E}_0 \times {\bf B}_0^*)+ c^2 \chi_{2, \rm BB}^{(\omega,-\omega)}({\bf B}_0^* \times {\bf B}_0 ) \Big],
    \end{aligned}
\end{equation}
where $\chi^{(\omega,-\omega)}_{2, \rm EE/EB/BB} = \sum_{\rm u= S, R}n_{\rm mol}^{(\rm u)}\beta^{\rm (u)}_{\rm EE/EB/BB}$ are the second-order NL-IFE susceptibilities. Importantly, we observe that, in spite of the random orientation, the CDS possesses second-order NL-IFE susceptibilities $\chi_{2,{\rm EE/EB/BB}}^{(\omega,-\omega)} = \chi_{2,{\rm EE/EB/BB}}^{(\omega,-\omega)}({\rm S}) + \chi_{2,{\rm EE/EB/BB}}^{(\omega,-\omega)}({\rm R})$, particularly chirally-insensitive scalar $\chi_{2, \rm EE}^{(\omega,-\omega)}\propto n_{\rm mol}$ and $\chi_{2, \rm BB}^{(\omega,-\omega)}\propto n_{\rm mol}$, as well as the {\it enantio-selective pseudoscalar} $\chi_{2,{\rm EB}}^{(\omega,-\omega)}\propto \Delta n_{\rm mol}$, which is proportional to the enantiomeric excess $\Delta n_{\rm mol} = n_{\rm mol}^{\rm (S)} - n_{\rm mol}^{\rm (R)}$. Figs.~\ref{Fig2}({\bf a-c}) depict the dependence of the complex NL-IFE susceptibilities $\chi_{2,\rm EE/BB/EB}^{(\omega,-\omega)}$ over the vacuum wavelength $\lambda$ for fixed $n_{\rm mol}^{\rm (S)} = 10^{-2}~{\rm nm}^{-3}$ of pure-S CDS, i.e., $ n_{\rm mol}^{\rm (R)} = 0$ and $n_{\rm mol}= n_{\rm mol}^{\rm (S)}$. Note that all the NL-IFE susceptibilities are resonant near the reparixin electronic resonance wavelength $\lambda \simeq 250~{\rm nm}$ (we emphasize that the broad resonant peak results from the convolution of 4 distinct narrower peaks arising from the transitions between the energy levels $\hbar\omega_i$ with $i=1-5$) and the chirally-insensitive contribution is mainly dominated by $\chi_{2, \rm EE}^{(\omega,-\omega)}$ since $\chi_{2, \rm BB}^{(\omega,-\omega)}$ is comparatively much weaker. In Fig.~\ref{Fig2}({\bf d}) we illustrate the dependence of $|\chi_{2,\rm EB}^{(\omega,-\omega)}|$ over $\lambda$ and total molecular number density $n_{\rm mol}=n_{\rm mol}^{\rm (S)}$ of pure-S CDS. Note that $\chi_{2,\rm EB}^{(\omega,-\omega)}\propto\Delta n_{\rm mol}$ flips sign for opposite enantiomerically pure CDSs and that in the optical spectral range $|\chi_{2,{\rm EB}}| \simeq 10^{-24} - 10^{-22}~{\rm m/V}$ for $\Delta n_{\rm mol} = n_{\rm mol} = 10^{-3}-10^{-1}~{\rm nm}^{-3}$, see Fig.~\ref{Fig2}({\bf d}). \\   

\begin{figure}[t!]
	\centering
	\begin{center}	
    \includegraphics[width=0.48\textwidth ]{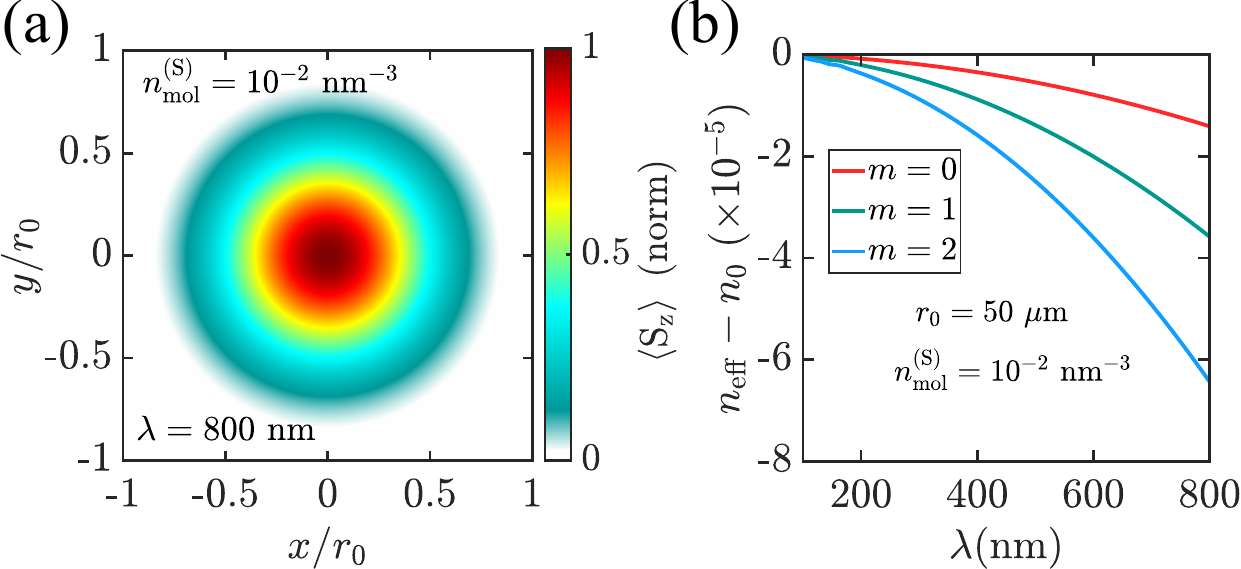}
		\caption{ (\textbf{a}) Dependence of the normalized time-averaged $z$-component of the Poynting vector $\langle S_z\rangle$ over normalized coordinates $x/r_0$ and $y/r_0$ perpendicular to the axis of the PMC with radius $r_0 = 50~\mu{\rm m}$, calculated at fixed impinging carrier wavelength $\lambda = 800~{\rm nm}$. (\textbf{b}) Dependence of the index correction $n_{\rm eff}(\lambda)-n_0(\lambda)]$ of the capillary over $\lambda$ for $n=1$ and diverse azimuthal indices $m = 0,1,2$. All results are obtained for a pure-S CDS with fixed $n_{\rm mol}=n_{\rm mol}^{\rm (S)} = 10^{-1}~{\rm nm}^{-3}$. } 
		\label{Fig3}
	\end{center}
\end{figure}

\begin{figure}[t!]
        \centering
        \includegraphics[width=0.49\textwidth]{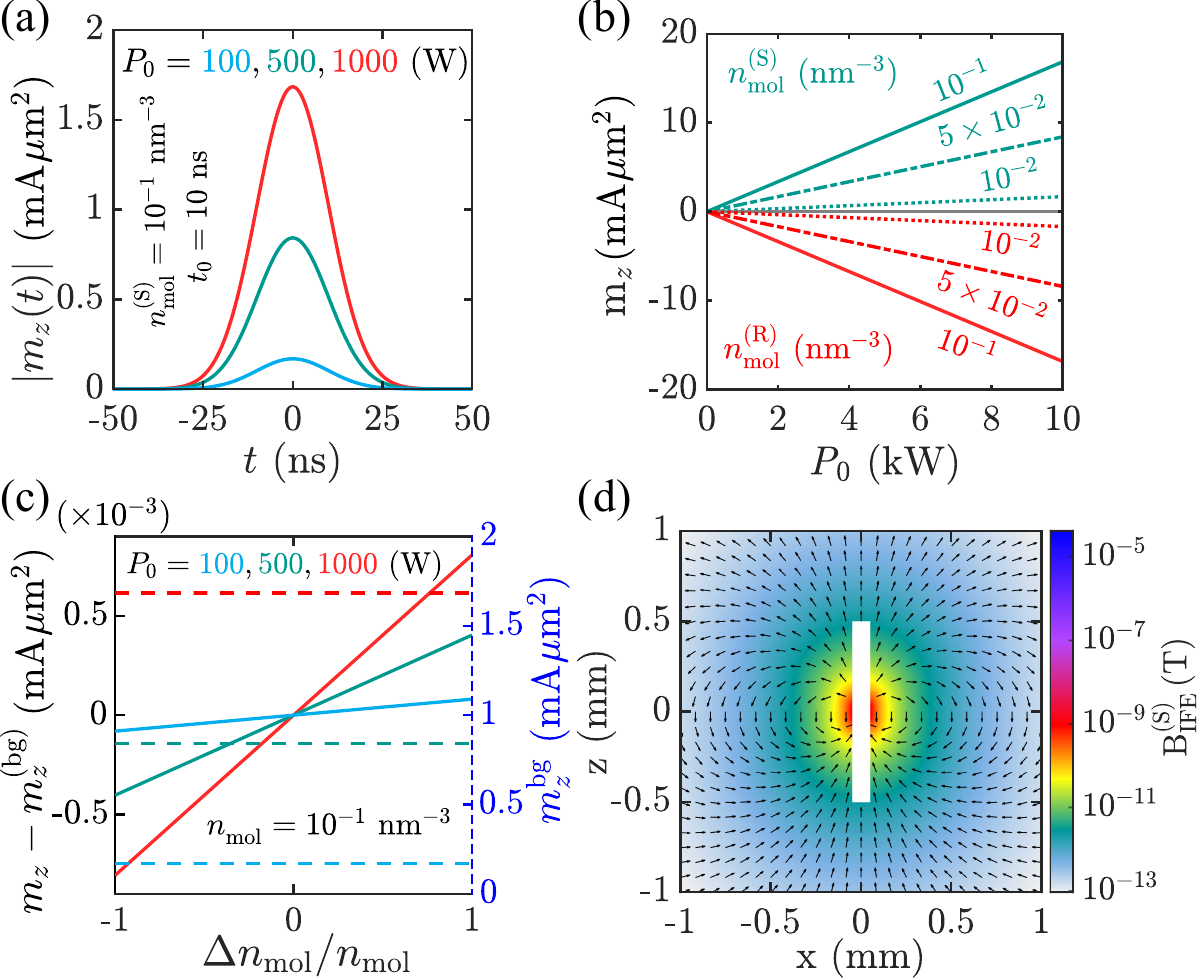}
        \caption{(\textbf{a}) Temporal evolution of IFE-induced magnetic dipole moment z-component modulus $|m_z(t)|$ generated in the considered PMC, see Fig.~\ref{Fig1}, for several distinct peak powers $P_0$ of the impinging laser with fixed pulse duration $t_0 = 10~{\rm ns}$ and right circular polarisation, calculated for pure-S CDS with $n_{\rm mol}^{\rm (S)}=10^{-1}~{\rm nm}^{-3}$. (\textbf{b}) Dependence of IFE-induced $m_z$ over $P_0$ for S/R CDSs with several distinct number densities $n_{\rm mol}^{\rm (S/R)}$. (\textbf{c}) Dependence of (left y-axis) chirally-sensitive $[m_z - m_z^{\rm (bg)}]$(solid lines) and (right y-axis) chirally-insensitive $m_z^{\rm (bg)}$ contributions to $m_z$ over the relative enantiomeric excess density $\Delta n_{\rm mol}/n_{\rm mol}$ for several $P_0$ and fixed total molecular number density $n_{\rm mol} = 10^{-1}~{\rm nm}^{-3}$. (\textbf{d}) Spatial dependence of IFE-induced magnetic field ${\bf B}^{\rm (S)}_{\rm IFE}({\bf r})$ over the normalized distances $x/l,z/l$ from the PMC axis, calculated for fixed $P_0 = 1~{\rm kW}$, embedding pure-S CDS with $n_{\rm mol}^{\rm (S)} = 10^{-1}~{\rm nm}^{-3}$, with the white shaded area at the center indicating the location of the PMC. All the results are calculated by considering length $l=1~{\rm mm}$ and radius $r_0 = 50~{\mu}{\rm m}$ of the PMC at fixed vacuum carrier wavelength $\lambda = 800~{\rm nm}$ of impinging radiation.} 
        \label{Fig4}
\end{figure}

\textit{IFE in PMCs embedding isotropic CDSs}--
In Fig.~\ref{Fig1}({\bf a}), we schematically illustrate the proposed device consisting of a glass capillary with radius $r_0=50~\mu{\rm m}$, length $l=1~{\rm mm}$, and embedding the CDS inside the core. The isotropic CDS dielectric response is characterised by the liner bi-anisotropic optical parameters, i.e., relative permittivity $\epsilon_{\rm r}(\lambda)$, relative permeability $\mu_{\rm r}(\lambda)$, and chiral parameter $\kappa (\lambda)$, see SI, and the NL-IFE magnetisation in Eq. (\ref{Eq5}). Note that the envisaged device may operate also through other photonic platforms, e.g., hollow core photonic crystal fibres, which have been  shown to produce ultrasensitive circular dichroism \cite{Helfrich2026}  Within the SVEA, we evaluate the electromagnetic excitation of the system upon impinging optical radiation at $z=0$ with carrier wavelength $\lambda$ and arbitrary polarisation defined by the superposition of right $(s=+1)$ and left $(s=-1)$ CP components. Assuming weak coupling between E and H modes by the CDS within the capillary core, we solve Maxwell’s equations in cylindrical coordinates $(r,\phi,z)$ for CP waves by projecting the incident electric/magnetic fields onto the impinging CP unit vectors $\hat{\bf e}_s^{(\pm)}=(1/\sqrt{2})(\hat{\bf e}_x \pm is \hat{\bf e}_y )$. We emphasize that we focus on the Marcatili approximation for capillary excitation, obtaining hybrid super modes
\begin{subequations} \label{Eq6}
\begin{align}
    &{\bf E}({\bf r},t) = \sum_{\alpha}{\rm Re} \left[A_{\alpha}({\bf r},t){\bf e}_{\alpha}(r) e^{i(\xi_{\alpha} -\omega t)} \right], \\
    &{\bf H}({\bf r},t) = \sum_{\alpha} \dfrac{1}{\mu_0 c} {\rm Re} \left[A_{\alpha}({\bf r},t){\bf h}_{\alpha}(r) e^{i(\xi_{\alpha} -\omega t)} \right], 
\end{align}
\end{subequations}
where we have introduced a single composite mode index $\alpha \equiv (n,m,s)$ such that $n$ and $m$ are the radial and azimuthal mode indices, respectively, $A_{\alpha}$ denote the mode envelopes, ${\bf e}_{\alpha}(r)$ and ${\bf h}_{\alpha}(r)$ are electric and magnetic vectorial mode profiles, explicitly provided in the SI. In addition, $\xi_{\alpha} (z,\phi) = (\beta_{n,m}+ [\Delta \beta_{n,m}]_s )z + m\phi$, where $\beta_{n,m}=\sqrt{(\omega^2/c^2){\rm Re}[\epsilon_{\rm r}\mu_{\rm r}]-{(1/r_0^2) \tilde{x}_n^2(m)}}$ is the unperturbed wavenumber, $[\Delta \beta_{n,m}]_{\pm}=[\omega^2/(2\beta_{n,m}c^2)](i \rm Im[\epsilon_{\rm r}\mu_{\rm r}]\mp2\kappa\sqrt{{\rm Re}[\epsilon_{\rm r}\mu_{\rm r}]})$  is the wavenumber correction, and $\tilde{x}_n(m)$ indicates the $n$-th root of the Bessel function $J_m[{\tilde{x}_n(m)}]$ of first kind of order $m$. In turn, we calculate the average Poynting vector $\langle \mathbf{S} \rangle = \left[ \mathbf{E}(\mathbf{r}) \times \mathbf{H}(\mathbf{r}) \right]$ for the mode propagation along the capillary. Fig.~\ref{Fig3}({\bf a}) illustrates the dependence of the calculated average Poynting vector projection $\langle S_z\rangle$ along the $z-$direction over the normalized coordinates $x/r_0$ and $y/r_0$ perpendicular to the capillary axis, providing the spatial distribution of the electromagnetic power flux within the capillary. In Fig.~\ref{Fig3}({\bf b}), we depict the dependence of refractive index contrast $[n_{\rm eff}(\lambda) - n_0(\lambda)]$, where $n_{\rm eff}=\beta_{n,m}/(2\pi)$ and $n_0 = \sqrt{\epsilon_{\rm r}\mu_{\rm r}}$, over the impinging wavelength $\lambda$ for different modes with fixed $n=1$ and $m=0,1,2$, calculated for pure-S CDS with $n_{\rm mol}^{\rm (S)} = 10^{-1}~ {\rm nm}^{-3}$ dissolved in MS. We excite the PMC by one such supermode, see Eq. (\ref{Eq6}), and calculate the corresponding NL-QSM, see Eq. (\ref{Eq5}), inside the capillary, which in turn, produces an IFE-induced magnetic dipole moment

\begin{align}\label{Eq7}
    &{\bf m} ({\bf r}) = \mathcal{F}(\omega) \Omega_{\alpha}(\omega) {\rm Re}[is\chi_{2,{\rm EE}}^{(\omega,-\omega)} + \zeta_s(\omega) \chi_{2,{\rm EB}}^{(\omega,-\omega)} ]\times, \nonumber \\
    &\quad \quad \quad \times |A_{\alpha}|\mathcal{J}[\tilde{x}_n(m)]\hat{e}_z,
\end{align}
where $\mathcal{F}(\omega) = 2\omega m^2 \beta_{n,m} r_0^4 \sqrt{{\rm Re}[\epsilon_{\rm r} \mu_{\rm r}]}/[\mu_0 c^2 r^2 \tilde{x}_n^4(m)]$, $\Omega_{\alpha}(\omega)= [e^{\vartheta_sl}-e^{-\vartheta_sl}]/\vartheta_s$, where $\vartheta_s ={\rm Im}[\Delta \beta_{n,m}^{(s)}] $, and $\zeta_s(\omega) = s\kappa^*+\mu_{\rm r}^*\sqrt{\rm Re[\epsilon_{\rm r} \mu_{\rm r}]}$, $\mathcal{J}[\tilde{x}_n(m)]=(1+J_0^2[\tilde{x}_n(m)]-2\sum_{k=0}^{m-1}J_l^2[\tilde{x}_n(m)])$. Note that while calculating the induced magnetic dipole moment, we neglect the contribution arising from $\chi_{2,{\rm BB}}$ since it contributes minimal corrections. Also note that ${\bf m} ({\bf r})$ contains chirally-insensitive contributions produced by $\chi_{2,{\rm EE}}^{(\omega,-\omega)}$, leading to background induced magnetic dipole moment
\begin{align}\label{Eq8}
    &{\bf m}^{\rm (bg)} ({\bf r}) = \mathcal{F}(\omega) \Omega_{\alpha}(\omega) {\rm Re}[is\chi_{2,{\rm EE}}^{(\omega,-\omega)}]|A_{\alpha}|\mathcal{J}[\tilde{x}_n(m)]\hat{e}_z,
\end{align}
along with the chirally-sensitive contribution assessed as $[{\bf m}({\bf r})-{\bf m}^{\rm (bg)}({\bf r})]\propto \Delta n_{\rm mol}$, governed by $\chi_{2,{\rm EB}}^{(\omega,-\omega)}$. By considering an input Gaussian laser pulse with intensity temporal profile $I(t)=I_0e^{-t^2/2t_0}$, where $I_0=(1/2)\epsilon_0c|\tilde{A}_0|^2$ is the peak intensity and $t_0$ is the pulse duration, we calculate such IFE-induced magnetic dipole moment. In Fig.~\ref{Fig4}({\bf a}), we report the temporal evolution of the induced magnetic dipole moment z-component modulus $|m_z(t)|$, calculated at fixed impinging vacuum wavelength $\lambda = 800~{\rm nm}$ and pulse duration $t_0 = 10~{\rm ns}$ for right $(s = +1)$ CP impinging radiation with several distinct peak powers $P_0  = 100, 500, 1000~{\rm W}$ with intensity normalized to the effective mode area of the capillary, embedding pure-S CDS with $n_{\rm mol}^{\rm (S)}= 10^{-1}~{\rm nm}^{-3}$. Fig.~\ref{Fig4}({\bf b}) illustrates the dependence of $m_z$ over $P_0$ at fixed $\lambda = 800~{\rm nm}$ for three distinct molecular densities of S/R-reparixin enantiomers dissolved in the MS. Note that the IFE-induced $m_z$ along the PMC axis increases linearly with $P_0$ and flips sign for opposite enantiomeric form. In Fig.~\ref{Fig4}({\bf c}), we report the dependence of (left $y$-axis) chirally-sensitive $[m_z-m_z^{\rm (bg)}]$ (solid lines) and (right $y$-axis) chirally-insensitive $m_z^{\rm (bg)}$  (dashed lines) contributions over the relative enantiomeric excess density $\Delta n_{\rm mol}/n_{\rm mol}$ for several $P_0$ and fixed total molecular number density $n_{\rm mol} = n_{\rm mol}^{\rm (S)} + n_{\rm mol}^{\rm (R)} = 10^{-1}~{\rm nm}^{-3}$. Note that the enantio-selective contribution to the IFE-induced magnetic dipole moment is sitting on top of a constant background induced magnetic dipole moment. Also note that $[ m_z-m_z^{\rm (bg)}]$ vanishes for racemic chiral mixtures and alters sign depending on the enantiomeric imbalance in the CDS. In turn, such an IFE-induced dipole moment ${\bf m}({\bf r})$, see Eq. (\ref{Eq8}), produces a QSMF in the vicinity of the PMC. In Fig.~\ref{Fig4}({\bf d}), we report the spatial distribution of the generated QSMF in the vicinity of the PMC with $l=1~{\rm mm}$ and radius $r_0 = 50~{\mu}{\rm m}$, calculated at fixed $\lambda = 800~{\rm nm}$ and $P_0 = 1~{\rm kW}$ for pure-S CDS with $n_{\rm mol}^{\rm (S)}= 10^{-1}~{\rm nm}^{-3}$. The arrows denote the direction (unit vector) of the generated QSMF, and the white shaded area at the center refers to the position of the PMC. Note that we obtain spatially varying IFE-induced QSMF of the order of $\simeq 10^{-8} - 10^{-13}$ T in the vicinity of the PMC. Indeed, the magnitude of such QSMF remains identical regardless of the enantiomeric form, while the direction of the QSMF flips sign for the opposite enantiomers. Thus, by measuring the IFE-induced QSMF, it is possible to retrieve the enantiomeric imbalance in the chiral mixture. \\

{\it Conclusion}--In conclusion, we theoretically demonstrate a novel chiroptical sensing paradigm capable of detecting enantiomeric imbalance in isotropic chiral mixtures with sample volumes as small as $\simeq 10$ nl. The proposed approach exploits a previously unexplored manifestation of the inverse Faraday effect in isotropic chiral media, whereby molecular chirality induces an effective second-order nonlinear magneto-optical response that survives orientational averaging. By combining first-principles quantum-chemical calculations with density-matrix modelling, we evaluate the molecular hyperpolarisabilities and derive the effective macroscopic nonlinear response of realistic chiral drug solutions within the electric- and magnetic-dipole approximation. We further design and benchmark a photonic micro-capillary platform operating with solvated reparixin, demonstrating the generation of measurable chirality-dependent quasi-static magnetic fields that directly encode the enantiomeric excess of the sample. These findings establish a fundamentally new route toward compact, highly sensitive, and integrable chiral sensing technologies, opening opportunities for lab-on-chip diagnostics, pharmaceutical quality control, and nonlinear nanophotonic devices for molecular enantiomer discrimination..

{\it Acknowledgements}--This work has been partially funded by the European Union - NextGenerationEU under the Italian Ministry of University and Research (MUR) National Innovation Ecosystem grant ECS00000041 - VITALITY -CUP E13C22001060006. This work has been supported by the European Union under grant agreement No 101046424. Views and opinions expressed are however those of the author(s) only and do not necessarily reflect those of the European Union or the European Innovation Council. Neither the European Union nor
the European Innovation Council can be held responsible for them.

\end{document}